\documentstyle[epsfig,subfigure,12pt]{article}
\def\he4{ ^4\!{\rm He} }
\def\c14{ ^{14}\!{\rm C} }

\def\be7{^7\!{\rm Be}}

\def\WAYNE{$^a$}

\def\PITT{$^b$}

\begin{document}
\title{Review of the Technical Issues Associated with the Construction of a Solar Neutrino TPC}
\author{ G. Bonvicini\WAYNE, 
         D. Naples\PITT, 
         V. Paolone\PITT
             \vspace{0.5cm} \\
\WAYNE{\it Wayne State University, Detroit MI 48201}\\
\PITT{\it Univ. of Pittsburgh, Pittsburgh PA 15260}\\
}
\maketitle

\begin{center}
{\bf Abstract}
\end{center}
\vskip .1in

In this note the issues surrounding the realization of a solar
neutrino TPC are reviewed. The technical challenges can be summarized as
building a very large TPC within the low background constraints of 
solar neutrino experimentation.

\newpage

\section{A comparative review of a solar neutrino TPC.}

A TPC is radically different from all other solar neutrino experiments in
that it measures the direction of the target particle (the electron). It also 
has a much finer granularity than other detectors, easily distinguishing 
between electron and non-electron events, and
events of multiplicity one and events of multiplicity two or higher.

Directionality provides much of the physics reach (by measuring the neutrino
energy and average flavor). From a feasibility point of view, however, we are
particularly interested in the effect that directionality has on background
suppression.
First, directionality allows a direct reduction of backgrounds by a factor of
four to ten. Only events pointing in a cone directly opposite the solar angle
are considered, with a $\cos\theta_{\odot}$ cut ranging from 0.5 to 0.8.
Second, directionality allows us to measure, up to statistical fluctuations,
all the backgrounds that affect the experiment. 
By ``all the backgrounds'' we mean both internal ($^{14}C$,
$^{85}Kr$, $Rn$) and external backgrounds ($^{238}U$, $^{232}Th$, $^{40}K$, and
cosmogenic activity in the surrounding materials).
Finally, since the target is mostly helium (we consider only 
He/CH$_4$ mixtures,
with helium at least 90\% of the mixture) internal cosmogenic activity
is virtually absent. This fact, plus the unique direct measurement of 
backgrounds translates into very modest depth requirements.

While low background experimentation applies stringent constraints to the
design of this detector, there are several aspects in which this TPC is
remarkably simple to build and operate, and these include most of the
major problems normally encountered by large TPC. The advantages are:
\begin{enumerate}
\item The TPC is located underground, inside a completely enclosed Faraday 
cage. We can reasonably assume that the noise seen by the TPC will be
generated only by its internal components. A complete list includes amplifier
noise, and capacitive noise from wires or strips plus cables. Both can be
reliably predicted at or below a few thousands electrons for an integration
time of order 250 nsec.
\item There are no constraints from other surrounding detectors. The HV can be
brought in with a very large cable (perhaps one foot in diameter), and the
HV degrader can be embedded in the 1-1.5 meter thick plastic shielding, 
providing
excellent electrostatic stability in the drift HV system.
\item The expected low rates ($\sim 0.1$Hz) virtually eliminate charge build-up
problems. One can design a detector plane where the wires are not directly
facing other electrodes, as described below, providing excellent electrostatic
stability in the detector plane.
\item The expected low rates virtually eliminate any chance of detector ageing,
allowing the option ot use materials such as uncoated steel wires or 
polyethylene, which would age a chamber rapidly in an accelerator environment.
\item A precision calibration is provided automatically by cosmic rays and
other background events, while calibration complexity is reduced by the lack
of a significant magnetic field.
\end{enumerate}

The TPC has one drawback; it is not self-shielded.
The region immediately surrounding the active volume contains significant
amounts of materials which are not as pure as the material in the fiducial
volume (the TPC cage, wires, strips, and the HV grid at least). This 
translates into substantial limitations on which materials can be
used in the surroundings. We will discuss the backgrounds, materials,
and their purities in Section 3.

While no TPCs of this scope have even been built, invaluable information
about the MUNU experiment was provided by C. Broggini\cite{broggini}. 
This is the largest underground pressure TPC built so far (about 2m$^3$), 
it collected data
on low-energy $\bar{\nu}-e$ scattering, and shared many of the problems
we will face in the near future. We refer to their experience throughout the
paper.

\section{The TPC as a Detector.}

\begin{figure}[pthb]
\centerline{\psfig{file=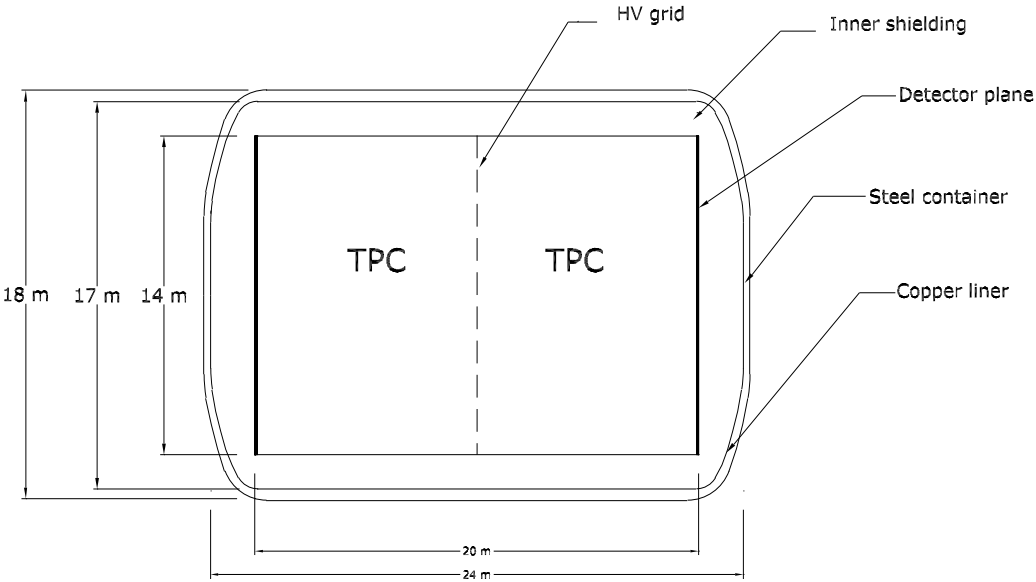,height=4.0in,width=5.5in,angle=0}}
     \caption{Sketch of TPC cage and surroundings.}
     \label{tpc}
\end{figure}

A sketch of the gross features of the proposed 
TPC is shown on Figure \ref{tpc}. It consists of a cylinder
twenty meters long and between 14 and 20 meters in diameter (depending on 
the final choice of detector parameters). It will contain between 7 and 10 Tons
of gas, and will be separated from the rock by 2.5 to 3.5 
meters of water equivalent 
of high purity shielding (in the figure, about 1.5 meters are provided by the
steel tank, and about 1.0 meters by high purity plastics/water). 
The mid plane is occupied by a high purity metal grid, at a voltage
of -100 to -200 kV. 
HV elements are recessed in the shielding. The TPC cylinder itself 
is enclosed
in an external pressure vessel. There is a slight overpressure inside the
TPC, compared to the rest of the pressure vessel. 
Teflon gaskets are used to seal 
the juncture of the barrel and endcap. 

The overall TPC design philosophy is one of a device that is as stripped down
as possible, minimizing the number of electronic channels and materials coming
into contact with the gas.
The end caps are the detector planes and contain a single set of wires and
a single set of strips, to be discussed in more detail later. 
The wires reconstruct the $(x,z)$ profile of a track, and the strips the
$(y,z)$ profile of a track, where $z$ is the drift direction. The
direction and energy of a track is reconstructed by combining both
track views. Assuming a pitch of 2 millimeters 
for wires and strips, the TPC needs between
28000 and 40000 electronic FADC channels. 

The gas mixture consists of boil-off helium in bulk, with a small admixture of
natural gas from a deep well boiled off at 112K (in practice, pure methane).
Boil-off at low temperatures eliminates all metals and many radioactive
or electronegative gases, from water vapor to radon.

Total gas volume and density determine the total number of target 
electrons for solar neutrinos. These are determined by the gross 
detector parameters; total drift length, wire/strip length, and gas pressure.
The total drift length is limited by diffusion considerations on the
drifting electrons. We have chosen a maximum drift length of ten
meters which corresponds to a maximum drift time of 
8.3~msec at 100 kV (measured drift velocity 1.2m/msec). The
total diffusion over ten meters drift for an
operating pressure of 10~atm is 4.3~mm
and changes only by a few percent at higher fields. 
The gas pressure is limited by a requirement on 
minimum track length to achieve our desired angular resolution,
and the wire length is limited by the RC constant being of order of the
FADC timing bucket (200-300 nanoseconds). The strip length is not significantly
limited at this point in time.

We will discuss the design of a TPC with 20m length, 14m diameter, 
2mm wire and strip segmentation, and 10 atm of He(97\%), CH4(3\%) in 
this document. We also assume for presentation purposes that the TPC
is located at a depth of 2500 mwe.

\subsection{Detector Plane and Grid}

Except for the central grid, which defines the HV midplane and will be
made of the cleanest metal wire available (probably OFHC
wire), the detector planes contain all the metal parts in the immediate
vicinity of the TPC.

\begin{figure}[pthb]
\centerline{\psfig{file=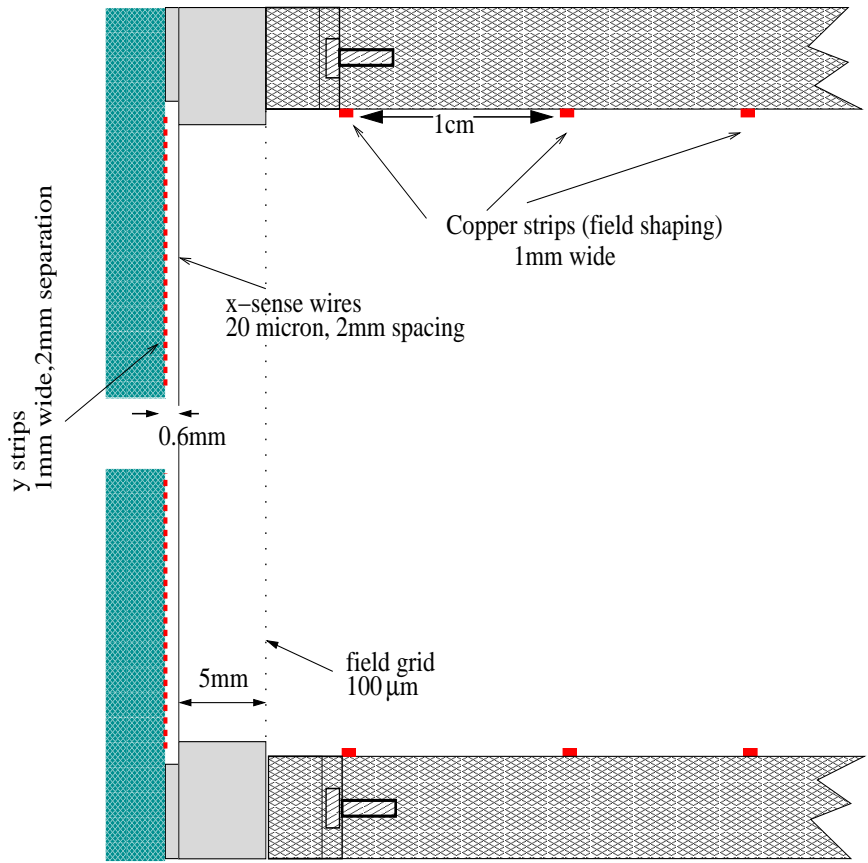,height=4.5in,width=4.5in}}
     \caption{Sketch of detector elements in TPC end cap planes.}
     \label{det_plane}
\end{figure}

The detector plane is shown schematically in Figure \ref{det_plane}. 
In its minimal 
configuration it consists of one plane of 20$\mu$m wires, and one plane
of strips, separated by about 0.6 millimeters, 0.1 millimeters of
which is in the form of a plastic foil insulating the two planes. The wires 
are held at a voltage of approximately 2-2.5 kV, 
and the strips are held at a negative voltage of about 1 kV with respect
to the wires, which forbids the presence of field lines breaking the wire
plane. There is a simple grid at 5 mm distance, separating the drift and 
gain region.

With the parameters discussed above, we expect a gain of $5\times 10^4$ to
$10^5$, with a pulse time constant of 200-300 nsec for the 
wires\cite{lpc9631}. By comparing to the expected electronic noise numbers,
we expect close to 100\% efficiency for the wires, 
while the coupling and efficiency for the strips is 60\% and 
80\%\cite{lpc9631}. The RC of the strips is much smaller than
that of the wires and depends on the thickness of the strip.

The wires can be made of any of the 
following three materials: steel, Cu/Be alloy, or uncoated tungsten. The 
latter is the material of choice for MUNU, and proved to be a radiopure
material by MUNU standards. The other two are choices based on the
known high purity of steel and copper. All three types of wires are 
readily available
commercially and need to be tested for radiopurity at a suitable facility.
The strips are made of high purity OFHC.

There is a premium in being able to assemble and disassemble the detector
plane and grid quickly: a quick assembly reduces initial 
contamination of surface
by radioactive dust or radon daughters, and a quick disassembly allows
to fix contamination problems by dust or radon, or to change the active
elements if they are found to be unsuitable. Therefore a TPC were all parts
are snap-on is considered.

The MUNU experience proves that the grid needs to be removable quickly. In one
large radon contamination episode (which was later tracked to the Oxysorb 
cartridge in use then), radon daughters, having been ionized in the nuclear 
decay, drifted as ions to the grid where they accumulated, probably by bonding
chemically to the metal (polonium has the same valence as oxygen). A circular 
curtain rail scheme, with a pulling cord and one-plug connection to the HV, 
should suffice.

The wires need to be strung without use of epoxy or solder. At this time a 
strategy is considered, where they are cut to measure and welded by 
electron gun to a small pin of the same material
at each end. The pin locks in an appropriate notch at the end plane rim.
In the case of tungsten wires,
the standard tension of 62 grams is equivalent to a 1\% elastic elongation. 
Therefore a 14 meters wire needs to be made 14 cm shorter if it is to be strung
at that tension, and that is what we intend to do. 
In fact, we consider a tension of 40g, to diminish the failure
rate, with the wires strung horizontally and supported by 1 millimeter thick
comb every 67 cm.

The combs, hung vertically, serve two purposes: they hold the wires, and 
they pin down the plastic sheet that separates electrostatically the wires
and strips. This sheet can be made of polyethylene, nylon, or kapton. In the
case of kapton, there exists the possibility of simply electrodepositing the
very long copper strips on one side, if radiopurity can be maintained 
through the deposition process. The strips RC constant can be controlled by
depositing a thicker layer of copper.
Another possibility is to have a clean plastic sheet, and a grooved
wall behind the sheet. The strips will then fit into the grooves, 
pinned at the top and hanging vertically.
The final design choice will depend 
on achievable radiopurity and ease of assembly.

\subsection{Gas System}

The special nature of this TPC places a particular importance
on this part of the apparatus. A gas system should basically consist of:
\begin{enumerate}
\item 0.05$\mu m$ dust filter
\item Oxysorb cartridge
\item Cold trap with charcoal filter operated just above the boiling 
point of methane, or 112K.
\end{enumerate}

The interaction between these parts of the gas system and further
issues of how they should be implemented are discussed more below.
For reference, the boiling point of methane is compared in Table 1 
with the boiling point of other relevant compounds. 
\begin{table}{htb}
\caption{Boiling points of gases of interest for the TPC.
\label{tn:Machines}}
\begin{center}
\begin{tabular}[t]{|c|c|}
\hline
  Gas   & Boiling Point (K) \\
 \hline
Methane             & 111        \\
Radon         & 211          \\
 Krypton  & 116  \\
water      & 373               \\
oxygen      & 87               \\
CF$_4$         & 145   \\
\hline
\end{tabular}
\end{center}
\end{table}


The main requirement on the gas mixture is radiopurity. 
The gas mixture can be obtained by mixing boil-off helium with deep well
natural gas boiled off at 112K, and by storing the mixture in a low-radon 
vessel and possibly underground. The mixture is non-toxic, non-explosive,
and non-flammable, minimizing hazards in case of a leak.
He/CH$_4$ has sufficient gain\cite{lpc9631},
 with wire HV at around 2.2 kV\cite{lpc9631} 
and low diffusion\cite{methmes,goro}. 

Electron lifetime for a drift of ten meters
distance has not been measured directly, however, there is 
substantial evidence that it can be achieved.
Icarus Collaboration has reported results for 1.4~m drift in 
liquid Argon,  a far more demanding medium for electron drift
with significant solubility for major electronegative
compounds (such as oxygen). \cite{icarus}. 
Also, 2.0 to 2.5 meters drifts have been obtained by 
major TPCs such as ALEPH and STAR, with
electron attenuation lengths exceeding 10 meters\cite{aleph,weiman}.

Considerable attention needs to be devoted to the gas impurities which
can degrade collection efficiency. In this subsection we discuss
those electronegative impurities (like water or oxygen) that can absorb
drifting electrons. The main detector component that can introduce
these impurities could be acrylic, which was found to be radiopure by
the SNO collaboration (who used it for their heavy water vessel) and then was
chosen by MUNU\cite{broggini} 
as the 
TPC cage material.
In the mixing process used in producing acrylic,
micro-bubbles of unmixed resin persist and become 
permanent production centers of both water and oxygen. 

To keep oxygen at workable levels, 
in MUNU, the gas was continuously recirculating through 
an Oxysorb,
\footnote{However, it is a possible risk that this could 
introduce radioactive impurities \cite{broggini}.}
with a recirculation time of 6 hours\cite{broggini}. 
In our case, recirculation time would be on the order of 
90 to 120 hours, if we assume the same typical
contamination and scale for the surface/volume ratio.
There are advantages in obtaining a much longer circulation time,
as discussed in the background section below.
Ideally, we would like to clean the gas from dust  
ONCE, clean the gas from oxygen as close as possible to ONCE, 
and have the cold trap on ALWAYS.

The gas system is also slightly complicated by having two circulation systems,
a helium-methane inside the balloon and another mixture (helium with
perhaps nitrogen, same proportions) outside the balloon and at a slightly
lower pressure. Helium moves freely across the balloon but the other gases
do not, and as a result helium is constantly leaking out of the TPC.

We are also considering polyethylene as an 
alternative to acrylic for TPC balloon material.
Polyethylene has the advantage that it
does not contain oxygen in the bulk material, and does not
produce oxygen containing compounds long term (once gases embedded
in the bulk material have diffused out).

We expect to have substantially less electronegative compounds
in the gas than in previous TPCs, for the following reasons in order of
increasing importance:
\begin{enumerate}
\item Our TPC will not contain any epoxies. These contribute
as additional source of both water and oxygen. 
\item Our gas will be continuously recirculated through a 
cold trap, at a temperature sufficient to remove both impurities 
as well as some radioactive gas contaminants.
\item Alternative balloon material, polyethylene, discussed above. 
\item Slight Helium overpressure will further inhibit the entry
of gases from the balloon from entering the active volume. 
\end{enumerate}

\subsection{Achievable TPC performance.}
100 keV tracks are considered the TPC ultimate benchmark. Such a track
at 10 Atm has a total length of 9 centimeters and generates 2500 electrons
that will drift to the anode while diffusing. Resolution effects are dominated
by angular resolution. The angular resolution, in turn, depends significantly
on multiple scattering, and the aspect ratio of the track (how long it is,
compared to how wide it is - basically, the ratio of track length versus
diffusion).

The average diffusion is equal to 2/3 of the maximum diffusion, or 2.8 
millimeters. 95\% of the electrons will be included within two standard
deviations in each direction. Therefore the typical 100 keV track will occupy
a volume of 1.1$\times$1.1$\times$9 cm$^3$ (the third component is along the
track direction). The TPC samples small unit volumes where two 
dimensions are given by the wire and strip pitch, and the third is equal
to the drift velocity times either the RC constant or FADC bucket (they
are about the same number). 

Thus, the sampled volume is approximately 
0.2$\times$0.2$\times$0.03 cm$^3$ (the third component is 
along the drift direction). There is an estimated  total of 
9000 samples in a 100 keV track, to be compared with the 
existing 2500 electrons
or 0.28 electrons per unit volume. In fact, the head of a track has a 
ionization
density substantially lower than the tail, bringing the ratio down to
close to 0.1 electrons per unit volume.

The estimate above shows that even a wire TPC recovers, for the most part,
single electron information\cite{goro} for tracks that have 
diffused considerably.
Tracks with less diffusion do not have single electron determination,
except in the tails of the diffusion distribution, 
but are also narrower and ultimately better determined.
We expect a somewhat slowly varying resolution across the TPC, where the
dilution due to diffusion is partially compensated by the turn-on of
single electron resolution. Note that measuring the two ($(x,z)$ and $(y,z)$)
track profiles is a factor $\sqrt{2}$ worse than determining each single
electron in $(x,y,z)$ - however such a loss is substantially smaller
than the gain of working at lower pressure compared to the original
design.

It is unclear, at this time, what is ultimately the average angular resolution,
what is the distribution of angular resolutions for a large number of tracks,
and how the resolution is related to $\chi^2$-like quantities that can be
extracted from the data themselves. The multiple-scattered track is in
fact a fractal of dimension two, and there is no book on how to fit (let alone
extract maximal information from) such a thing. It will take substantial 
software development to arrive at a final answer. 

We wish to point out at this time
that, if extra resolution is needed to effectively cover the physics, 
we can always build a lower pressure, larger TPC. We estimate 
conservatively a $P^{-2}$ 
dependence of the resolution on pressure.
From our simulations we estimate a resolution of 15 degrees at 100 keV and
10 Atm, and decreasing like $T^{-0.6}$, $T$ being the electron kinetic 
energy.

\subsection{TPC Calibration.}
We have touched on calibration issues in the Physics 
Reach Paper\cite{snowmass} (attached).
Clearly identified background events provide gold plated calibration
events. Assuming a depth at which the cosmic rate through the chamber
is 0.1Hz, and where the bulk purity of the materials surrounding the TPC
is 10$^{-12}$g/g of $^{238}U$, yearly rates of interesting events are as 
follows: 
\begin{enumerate}
\item 3$\times$10$^6$ cosmic rays providing wire and strip gain calibration.
\item 7$\times$10$^5$ cosmic rays crossing the grid or one of the planes
providing diffusion and drift velocity calibration.
\item 3$\times$10$^5$ cosmic rays crossing both the grid or one of the planes,
providing drift and diffusion calibration over the whole drift.
\item 3$\times$10$^5$ delta rays with $T>100$ keV, where angle and
kinetic energy are
kinematically related.
\item 1$\times$10$^4$ double Compton scattering, where events of multiplicity
two are observed and are correlated in angle. Several kinematic relations
can be used to constrain the angular resolution and energy resolution
at the 1\% level.
\end{enumerate}
We expect no significant error due to incomplete
knowledge of the TPC performance.

\subsection{Pressure Vessel \& Engineering}
The pressure vessel consists of a copper or steel 
liner which is airtight and pushes
against the steel tank. The copper liner would be assembled in situ out of
large copper sheets which are welded together. Defects would be seen by using
Thermal Wave Imaging (TWI), a technique that can easily find 10 microns 
hairline cracks or less in the field. 

A first test of copper welded plates against
TWI was performed at WSU and was completely successful - areas where the 
welding was incomplete
were clearly seen, while areas with good welding tested positive. We have
a CD with six TWI movies that will be made available on the web within the 
next month. Currently, steel sheets and OFHC sheets are being tested at WSU
to decide which liner is best suited to the pressure vessel.

Once the pressure liner is completed, it becomes a high class clean room
while the detector is built from the inside out. Because the steel tank
is by far the strongest part of the device, there might be a need to have
large rods, made of the same material as the liner, 
mechanically connected to the tank, welded to the liner, and available to 
support parts inside of the liner. 

In principle, the plastic
shielding of Figure \ref{tpc} is assembled via roman arch assembly 
and any extra
support, if needed, is provided by the rods. The TPC cage is supported
by the plastic shielding. The total pressure due to the strung wires
is a modest 280 kg per side, whereas the total weight of the plastic
shield is 1400 Tons.

\section{Backgrounds.}
The signal sample is defined as contained events of 
multiplicity one. In addition these events must have 
clearly elongated tracks, 
and a dE/dx profile compatible
with the electron hypothesis. 
These simple requirements eliminate the backgrounds due to
$\alpha$'s, neutrons, cosmic rays, and $\beta$ radiation entering through
the TPC barrel. 

The two classes of events that remain as background are single 
Compton electrons,
regardless of the $\gamma$ source, and $\beta$-decay within the active volume.
Some sources, that do not produce backgrounds per se but could generate
unacceptable noise levels, are also discussed in the following sections.

$\gamma$ rays are produced predominantly in the radioactive decay of $^{238}U$,
 $^{232}Th$,  and $^{40}K$, with roughly 2, 2 and 1 gamma's of suitable 
energy produced in each completed chain respectively. The $^{238}U$ $\gamma$
spectrum
is shown in Fig. 3 of Ref.\cite{kessler}. 

$^{238}U$ and $^{232}Th$ also
produce a small amount of 
fission neutrons, which do not produce a background directly but
are not attenuated much by high Z materials and generate penetrating $\gamma$
of several MeV on being captured. The effective rate of neutrons does depend
on the exact composition of the producing material, however, it is typically
down by a factor of 1000 compared to the $\gamma$ rate.

In the accompanying paper\cite{snowmass}, based on the 
analysis by R. Kessler\cite{kessler}, it was found that a total 
background rate of about 150 events per day above 100 keV and 
before directional cuts
%
generate a small background subtraction error 
over one year (and negligible over the lifetime of the experiment). To achieve
this rate we must maintain a limit of $\sim 0.5\mu$grams of $^{238}$U 
equivalent, uniformly 
distributed on the inside wall of the TPC cage, plus a 
maximum $^{14}$C to $^{12}C$
ratio of $10^{-19}$ in the TPC gas. 

$^{14}$C, however, is a $\beta$ emitter and in general can only 
generate a background
if it forms part of the TPC gas mixture or the electron produced enters the 
active TPC region through the detector endcaps.

In this Section, we work our way to Table 2, that characterizes the background
levels that need to be obtained for this TPC to produce physics results.
We consider seven sources of backgrounds, and for each source, the purity that
will provide a non-dominant
background subtraction error after six months of data taking. In practice,
we quote the purity levels that will produce 150 electron tracks per day,
with energy greater 
than 100 keV, but before any directional analysis.
Subtraction errors add
quadratically, so if any four sources considered are at the value quoted,
the background subtraction error would become non-dominant after one year.
We make extensive use of the reduction factors calculated by R. Kessler
elsewhere\cite{kessler}.

\subsection{Cosmogenic backgrounds.}
Regardless of depth, the dominating cosmogenic background for this TPC is the
generation of $^{11}C$(a positron emitter) in the inner plastic 
shield and methane. The only
radioactive daughter from cosmogenesis in helium is tritium, which has an
end-point of 17.6 keV, well below our 100 keV threshold. Cosmogenesis 
in the iron shell
is also quite small, with no photons produced when a proton, neutron or
$\alpha$ is extracted from $^{56}Fe$. 

In practice, the limiting depth factor is not so much the irreducible
Compton backgrounds but rather the advantage of keeping the cosmic 
rate at or below 0.1 Hz (this corresponds to a depth of 2500 mwe), to minimize
unnecessary vetos.

This class of backgrounds, at a depth of 2500 mwe, contributes less than 
ten events per 
day mostly due to the generation of $^{11}C$ in the inner plastic 
shield, and less in 
the methane. It is considered negligible and subsequently ignored.

\subsection{Backgrounds from Surrounding Rock.}
Simulations indicate that $\gamma$ radiation from surrounding rock can be
attenuated to negligible levels (ten events per day or less) with reasonable
shielding\cite{arza}. For a rock purity of 5X10$^{-8}$g/g (WIPP site\cite{wipp}), 
3 mwe is sufficient. The 
Homestake site (rock purity 3X10$^{-6}$g/g \cite{homestake}), 
will require an additional 1 mwe (4 mwe total) shielding.
In our case most of the shielding is provided by the iron shell, which is required to be
30cm thick at Homestake and 20cm at WIPP. 

Iron does not appreciably attenuate low energy neutrons\cite{arza},
which have a flux about 1000 times lower than the $\gamma$.
This source can be eliminated (less than one event per day) if
a 5 cm plastic liner is inserted between the iron and the surrounding rock 
cavern (or, following the MUNU example\cite{broggini}, 
with 1 cm of boron-doped polyethylene).
The plastic effectively stops the neutrons via the reaction $^1H(n,\gamma)d$
(if it is boron-doped, no photons will be emitted).
The produced 2.2 MeV $\gamma$, if directed toward the active TPC region, is 
mostly absorbed
by the iron shell and inner plastic shielding.

\subsection{Backgrounds($\gamma$'s) Generated from within the Iron Shell.}
If the overall steel purity is $10^{-11}$g/g the 
amount of $^{238}$U uniformly distributed throughout the iron shell volume 
is 0.04 grams.
A 30 cm iron shell would effectly reduce this amount to 
0.7 milligrams of $^{238}$U 
placed uniformly at the surface of the iron/inner plastic interface. 
This amount is further attenuated by a factor of $10^3$ by 100 cm of plastic
before reaching the active TPC region. Therefore the effective contribution of this
uranium impurity is equivalent to placing 0.7$\mu$grams uniformly 
distributed on the inside surface of the TPC cage which generates $\sim 200$ background
events/day before directional analysis. In practice the photons which reach
the TPC have a severely degraded spectrum which further reduce the probability
of generating 100 keV electrons by about a factor of three, or 0.23$\mu$grams
total.
%

\subsection{Backgrounds Generated by the TPC Cage, Detector Planes and Inner Plastic Shielding.}
A 10$^{-12}$g/g contamination of $^{238}U$ in the plastic 
(a purity similar to the limit obtained by SNO\cite{SNO})
corresponds
to 1.3 milligrams of $^{238}U$ uniformly distributed throughout the inner
plastic shielding and TPC cage. The plastic self-shielding is
equivalent to placing 50$\mu$grams uniformly on the inside surface of the 
TPC cage. With some reduction factor due to energy degradation, there will
be $\sim 15000$ background events/day before directional analysis.

Clearly there is a need to go beyond the limits set by 
SNO\cite{SNO}, by two orders of magnitude, if this
source is to be brought down to the level of the other sources. 
One possibility, of course, is that the true value is not just below
the SNO limit.
Another  
possibility is to have the last part of the plastic shield in the form
of a deionized water vessel. Deionized water is known to have a radioactive
contamination not exceeding $10^{-14}$g/g. Clearly a major effort of
the RD program is to find a viable solution to this problem. 

Because of the much larger mass of the plastics compared to the active 
elements of the TPC (wires, strips, and grid, which are assumed to have
a total weight of 30 kg) a purity of 
3X10$^{-11}$ g/g is needed for
these elements, which is above the currently measured 
purity limits for OFHC and some steels ($10^{-11}$g/g). 


\subsection{$^{85}$Kr Contamination.}
$^{85}$Kr, which is a $\beta-$emitter, is only a background in 
our detector if it is introduced into our 
helium/methane gas mixture.
$^{85}$Kr was found to be a significant source of background by the Borexino
Collaboration\cite{borexino}, is produced via nuclear fission of 
Uranium and 
Thorium, and has a half-life of 10.8 years.
In an underground tunnel the air
contains some $^{85}$Kr and, in the case of Borexino, enough to contaminate
their open air water shield. 

It is unlikely that $^{85}$Kr will be a problem
for our TPC. Exposure to the underground air will be much less than that
at the Borexino Counting Testing Facility, and at startup the TPC gas will
be changed and vented  several times before recirculation is started. 
This process should reduce the concentration of $^{85}$Kr to negligible levels.
We can also expect reductions from our cold trap as Krypton is liquid at 112K
(see Table 1). The level of $^{85}$Kr activity in our gas mixture 
must be kept below 0.7$\mu$Bq/m$^3$($\sim$150 events/day).

\subsection{Radon Contamination.}
The effect of radon contamination in our gas mixture is largely
reducible but must be
kept at or 
below 0.7$\mu$Bq/m$^3$($\sim$ 150 events/day before directional cuts). 
The Radon decay chain reads as follows (half lives are in parentheses):
\[^{222}Rn(3.8 d) \to ^{218}Po(3.1m) +\alpha \to ^{214}Pb(27m) +\alpha\\
\to ^{214}Bi(20m) +\beta+\gamma \to ^{214}Po(160\mu sec) +\beta+\gamma \]
\[^{214}Po \to ^{210}Pb(22y) +\alpha.\]
There are no appreciable backgrounds beyond $^{210}Pb$ for a 
chamber service period of order one year. The $^{214}Bi$ leg of the decay
chain is reduced using the signature of a single electron followed by
an $\alpha$ from the $^{214}Po$ decay within a 1/2 millisecond window.
The predominant irreducible background comes from the $^{214}Pb$ beta decay.

The effect of the radon decay chain can be significantly reduced. 
In the presence of an electric field, the ion $^{218}Po$, with a lifetime
of a few minutes, drifts to the
HV grid where it can bond chemically with the metal. In case of large radon 
contamination, the grid will itself become contaminated, however, it will
also act as a radioactive trap, effectively segregating most of the $^{210}Pb$ and
localizing the background level in z. Our TPC will be designed with a 
removable HV grid for easy service.
%

Radon was the dominant background for the MUNU\cite{broggini} collaboration.
Initially it was an unacceptably large source of background, 
until it was traced to an Oxysorb cartridge. These are chemically
active filters which remove many compounds from a gas mixture. The cartridges 
are commercially made with
little possibility of controlling their radiopurity. 
One of the goals of the R\&D program is to explore all avenues that
could lead to an Oxysorb-free gas circulation system.

After the MUNU collaboration changed its
Oxysorb, the radon level stabilized to 10$^{-3}$Bq/m$^3$,
see figure 14 in reference\cite{broggini}.
This value however is still three orders of
magnitude above what we require.

There are three facts that point to a much smaller radon concentration in
our TPC compared to MUNU. They are, in increasing order of importance:
\begin{enumerate}
\item A much larger volume to surface ratio, not only in the TPC but also
between the TPC and the gas system, which should produce significant radon dilution (at least a factor of ten). The dilution factor should hold also when
comparing this device to the GNO experiment discussed below. 
\item  With a slight overpressure inside the TPC cage, outgassing from
the cage should be eliminated. This would greatly reduce
radon levels, if the radon source is in the TPC cage. It will almost certainly
suppress oxygen levels in the TPC gas and in turn 
possibly eliminate the need for an Oxysorb in the gas circuit.
\item A cold trap with a charcoal filter, similar to what has been
implemented in the GNO experiment will be used. 
GNO achieved concentrations of 0.2$\mu$Bq/m$^3$\cite{heusser}. 
However our cold trap will operate at 112K, as
opposed to the 77K used in GNO. The loss in radon capturing efficiency, due to the 
higher temperature, should be of order two or three\cite{heusser}. 
\end{enumerate}

\subsection{Dust Contamination.}
Dust entering the TPC active region will be of radiopurity similar to that of
the surrounding rock and deposit itself on the inside wall of the TPC cage.  
However many low-background experiments have proven that dust contamination of 
surfaces can be  greatly reduced by thorough
cleaning with deionized water or distilled alcohol. The dust budget of our
TPC, is ten grams at WIPP and 100
milligrams at Homestake($\sim$ 150 decays/day).
%
%
Dust usually has a fairly invariant size distribution, peaking at or around
1 micron diameter. While larger dust will settle over days, sub-micron
dust particles can remain airborne for weeks or months. A 0.05 micron dust 
filter should be able to remove more than 99\% of all airborne dust. We have no
data about the radiopurity of a dust filter, however such a filter has been 
used by the Borexino collaboration without any negative effects. We plan to insert
a dust filter in the gas system only after each TPC access and then remove it
when the TPC is in operation. 

\subsection{$^{14}$C Contamination in the TPC Active Region}

A substantial $^{14}$C contamination does not eliminate a measurement
of the all-important $(pp)$ neutrino flux. The electron produced in the $^{14}$C
decay has an end-point energy of 156 keV. For $(pp)$ neutrinos the 
maximum electron recoil energy is
217 keV.
However observables sensitive to neutrino
mixing, such as the electron recoil spectrum, and a distortion of
the $(pp)$ energy spectrum, would suffer since we would be sensitive to
a smaller fraction of the total $(pp)$ flux. There is no doubt that a high  
$^{14}$C content 
would significantly affect the physics reach of the detector and
therefore must be reduced to an acceptable level.

Our accompanying paper\cite{snowmass} describes a simulation where a  
$^{14}$C$/^{12}$C ratio of $\sim$5X10$^{-20}$ was assumed which generates
$\sim$20 background events/day before directional cuts. 
We scale that result to find that the limit here (for 150 events/day) is
$\sim$3X10$^{-19}$.
A similar carbon purity requirement was found, for the case of CF$_4$ as a TPC gas,
by Arpesella and Broggini\cite{arpe}.

\subsubsection{$^{14}$C Contamination in Methane Gas(CH$_4$).}

The need for quenching requires the use of some
fraction of methane in our Helium gas. Deep-well hydrocarbons 
are supposed to have maximal purity, with an 
estimated $^{14}$C$/^{12}$C ratio of 5X10$^{-21}$\cite{borc14}. Sources 
of $^{14}$C consist mostly of nuclear interactions in the rock, such as the reaction
$^{17}$O(n,$\alpha$)$^{14}$C\cite{borc14}. The created $^{14}$C is then 
assumed to form (or attach itself to) an hydrocarbon. 
Creation of $^{14}$C in the 
petroleum or gas directly is due to neutron capture by $^{13}$C, however it
is considered only the third most important creation process\cite{borc14}.
The source of $\alpha$'s and neutrons for
such reactions to take place are generated from the fission of uranium and
thorium in the surrounding rock. Consequently the abundance of these elements
in the adjacent rock should determine the level of $^{14}$C. It is therefore
natural to assume that the $^{14}$C$/^{12}$C ratio in deep-well methane
should vary from site to site. 

In addition the $^{14}$C content of methane should
be significantly lower than that of petroleum or more 
complex hydrocarbon gases. 
The created $^{14}$C will most likely react chemically with an
existing hydrocarbon and not with hydrogen molecules which are
very scarce. This chemical process will create molecules with at least two 
carbon atoms which can be separated from methane via boil-off.
We are currently trying to establish the chemical history of single carbon
atoms in a deep-well environment to validate this hypothesis and discussions
with geochemists have been started.

If our hypothesis is correct, then the dominant source 
of $^{14}$C in methane is from neutron capture 
by $^{13}$C. This implies that the $^{14}$C content of 
methane could be highly variable. 
The level would depend not only on uranium concentration, 
but also on boron concentration. Further, the chemical history
of the created $^{14}$C is controlled via the chemical potentials by the
local temperature and
pressure.

The direct Borexino measurement of the $^{14}$C content in their 
liquid scintillator\cite{borc14} was found
to be a factor of 190 higher than our requirement, or 1.9X10$^{-18}$g/g.
Originally explained as contamination by atmospheric carbon in the 
petrochemical plant, the problem has been reanalyzed by Schoenert
\cite{schoenert}. According to Schoenert, there exists the possibility that
the Borexino measurement does reflect the true content of  $^{14}$C
in petroleum. 

Even if the Borexino experiment measured the
true content of $^{14}$C, it did so only for the complex 
hydrocarbons(molecules
with more that one carbon atom) that exist in their liquid scintillator.
In fact a measurement of the $^{14}$C content in deep-well methane was
performed\cite{ams} and an upper limit of 1.6$\times10^{-18}$ was obtained.
Even though we require a limit of one order of magnitude better, this result
along with the Borexino measurement is consistent with our hypothesis.   

If our hypothesis about the chemical history of single carbon is valid, 
it is possible that only  
methane or compounds synthesized from methane should be used. 
A goal of the R\&D program is to determine adequate sources of methane that
meet these requirements.  

\subsubsection{$^{14}$C in the barrel.}
Although it does not present a significant source of background, there
will be $^{14}$C in the TPC cage. The cage is separated into a
barrel and two endcaps and will be discussed separately.
$^{14}$C decays in the barrel will result in tracks entering the active
volume. While they will be easily eliminated by the selection cuts, 
it is important that the rate be low enough that such events do not
generate unnecessary vetoes. In fact, if the results from Borexino are used,
less than one hundred trackable electrons per day will enter the 
TPC, generating a negligible vetoing rate.
$^{14}$C decays will also generate $\gamma$ rays via internal or external
bremmstrahlung (this is true both for the barrel or the endcaps). In either 
case, the maximum energy of the $\gamma$ is 156 keV, which can generate a maximum
electron recoil energy of 58 keV, well below our threshold. In addition both 
bremmstrahlungs are suppressed, compared to the rate of electrons entering
the TPC, by at least one order of magnitude.

\subsubsection{$^{14}$C in the endcaps.}

The fundamental difference between $^{14}$C in the endcaps and the barrel is that, 
if an electron track emanates from the endcap, it will be considered a candidate. These
electrons have no clearly reconstructed entry point in any material. If the Borexino 
measured $^{14}$C content applies to all
petroleum derivatives this background is not a problem.
In such a case, there will be few tens of background events per day of this type. 

Further cuts can only strongly suppress this background. The
diffusion perpendicular to the track direction will be well measured, and that in turn 
measures the $z-$coordinate along the drift direction. The 
$z-$resolution is of order one meter for ten meters drift, but is of
order a few centimeters
near the detector plane. Such cuts would eliminate negligible amounts of signal
events (by effectively shortening the active volume by a few centimeters).

\subsection{Background Summary}
A summary of estimated background rates, before signal cuts, from the the different 
sources is presented in Table \ref{bkgd_sum}. It's obvious from the the above
discussions that the R\&D program involving background reductions 
will involve determining adequate sources of methane, the design of a gas 
circulation system that reduces radioactive contaminations while 
not introducing new ones and an effective plastic/water inner shield.

\begin{table}[pbht]
\begin{center}
\begin{tabular}{||l|l||}
\hline\hline
Source & Requirements \\ \hline\hline
Cosmogenesis & $>$1800 mwe overburden \\ 
Rock $\gamma$ & 3-4 mwe total shielding \\ 
Rock neutron & plastic liner \\ 
Iron Shell & 1.5 mwe inner shielding \\
&(with $\sim10^{-11}$g/g)\\ 
Inner Plastic Shielding and TPC Cage  & 3X$10^{-13}$g/g\\ 
Detector Planes & 3X$10^{-11}$g/g\\\ 
$^{85}Kr$ in TPC Gas & 0.7$\mu$Bq/m$^3$\\
Radon in TPC Gas & 0.7$\mu$Bq/m$^3$\\
Dust on TPC Cage& 100 mg - 10g\\
$^{14}C$ in TPC Gas & $3X10^{-19}$ $^{14}C/^{12}C$ \\
$^{14}C$ in TPC Cage & $10^{-17}$ $^{14}C/^{12}C$ \\
\hline\hline
\end{tabular}
\end{center}
\caption{Summary of purity/shielding requirements. Each number refers to
the level at which the background subtraction error becomes non-dominant
after six months of data taking. If two numbers are quoted, the first refers
to Homestake, and the second to WIPP.}
\label{bkgd_sum}
\end{table}

\end{document}